\documentclass[aps,eqsecnum,twocolumn]{revtex4}
\usepackage{epsfig}
\newcommand{\beq}{\begin{equation}}
\newcommand{\eeq}{\end{equation}}
\newcommand{\beqs}{\begin{eqnarray}}
\newcommand{\eeqs}{\end{eqnarray}}

\newcommand{\lsim}{\mathrel{\raisebox{-.6ex}{$\stackrel{\textstyle<}{\sim}$}}}
\newcommand{\gsim}{\mathrel{\raisebox{-.6ex}{$\stackrel{\textstyle>}{\sim}$}}}

\begin{document}

\title{Exact Results for Average Cluster Numbers in Bond Percolation on 
Lattice Strips}

\author{Shu-Chiuan Chang}

\affiliation{Physics Division \\
National Center for Theoretical Sciences at Taipei \\
National Taiwan University  \\
Taipei 10617, Taiwan}

\author{Robert Shrock}

\affiliation{C. N. Yang Institute for Theoretical Physics \\
State University of New York \\
Stony Brook, N. Y. 11794}

\baselineskip 4.5mm

\begin{abstract}

We present exact calculations of the average number of connected clusters per
site, $\langle k \rangle$, as a function of bond occupation probability $p$,
for the bond percolation problem on infinite-length strips of finite width
$L_y$, of the square, triangular, honeycomb, and kagom\'e lattices $\Lambda$
with various boundary conditions.  These are used to study the approach of
$\langle k \rangle$, for a given $p$ and $\Lambda$, to its value on the
two-dimensional lattice as the strip width increases.  We investigate the
singularities of $\langle k \rangle$ in the complex $p$ plane and their
influence on the radii of convergence of the Taylor series expansions of
$\langle k \rangle$ about $p=0$ and $p=1$. 

\end{abstract}


\maketitle

\section{Introduction}

The study of percolation gives insight into a number of important phenomena
such as the passage of fluids through porous media and the effect of lattice
defects and disorder on critical phenomena.  Here we consider bond percolation.
Let $G=G(V,E)$ be a connected graph defined by a set $V$ of vertices (sites)
and a set $E$ of edges (bonds) connecting pairs of vertices.  Multiple bonds
connecting the same pair of vertices are allowed, although most of the graphs
that we study have only simple bonds.  We denote the number of vertices and
bonds as $n=n(G)=|V|$ and $e(G)=|E|$.  One now envisions generating a large
ensemble of corresponding graphs in which each bond is randomly present with
probability $p \in [0,1]$.  In the usual statistical mechanics context, one is
interested in a $d$-dimensional thermodynamic limit of a regular lattice graph 
$\Lambda$ in which the bonds are present with probability $p$.  Consider the
probability $P(\Lambda,p)$ that a given site belongs to an infinite cluster.
For a given lattice $\Lambda$, as $p$ decreases from 1, $P(\Lambda,p)$
decreases monotonically until, at a critical value, $p_{c,\Lambda}$, it
vanishes and remains identically zero for $0 \le p \le p_{c,\Lambda}$.  This
has an important effect on physical phenomena that take place on such a
bond-diluted lattice.  For example, consider a spin-model (with finite-range
spin-spin interactions) on $\Lambda$ and assume that for the undiluted lattice
the model is above its lower critical dimensionality so that there is a phase
transition at a finite temperature $T_c$.  Then as $p$ decreases below unity,
$T_c(p)$ decreases below $T_c(1)$, and as $p$ decreases to $p_{c,\Lambda}$,
$T_c(p)$ vanishes nonanalytically and remains zero for $0 \le p \le
p_{c,\Lambda}$. Other quantities also behave nonanalytically at
$p_{c,\Lambda}$, such as the mean size $S(\Lambda,p)$ of the percolation
cluster, which diverges as $p$ increases through $p_{c,\Lambda}$. Some previous
literature on bond percolation relevant to our present work includes
Refs. \cite{hammersley}-\cite{counter}.

An interesting quantity in this context is the number of connected components
(clusters), including single sites, for a given lattice $\Lambda$, divided by
the number of sites on the lattice and averaged over all of the graphs in the
above ensemble.  We denote this mean cluster number per site as $\langle k
\rangle_\Lambda$. Explicit exact values for $\langle k \rangle_\Lambda$ at the
respective values $p=p_{c,\Lambda}$, denoted $\langle k \rangle_{c,\Lambda}$,
have been calculated in Ref. \cite{ziff97} for the square, triangular, and
honeycomb lattices (where they were denoted $n_c^{B-\Lambda}$). However, to our
knowledge, $\langle k \rangle$ has never been calculated exactly as a function
of $p$ for the full range $p \in [0,1]$ for any lattice except a
one-dimensional or Bethe lattice \cite{bethe,stanley77,stauffer}.

In this paper we present exact calculations of this average cluster number per
site, $\langle k \rangle$, as a function of $p$, for a variety of
infinite-length, finite-width strips of regular lattices. We consider strips of
the square, triangular, honeycomb, and kagom\'e lattices.  These are of
interest since, at least for modest strip widths, one can obtain explicit
analytic expressions for $\langle k \rangle$ and can exactly determine, e.g.,
singularities that these expressions have in the complex $p$ plane and their
influence on series expansions.  When referring to a specific lattice strip
$\Lambda_s$ (including transverse boundary conditions), we shall denote the
average cluster number per site as $\langle k \rangle_{\Lambda_s}$. 
Our results interpolate between the known exact solutions for the
one-dimensional lattice (line) and the case of two dimensions, and complement
numerical simulations and series expansions.  Early studies of percolation on
infinite-length, finite-width strips include
Refs. \cite{derrida1,derridastauffer}, which focused on universal properties
such as critical exponents.  While the average cluster number is obviously
nonuniversal, depending as it does on the specific type of lattice, it still
give useful information about the phenomenon of percolation, as we shall show.

We take the longitudinal and transverse directions to be $x$ and $y$ and denote
the size of the lattice strips in these directions as $L_x$ and $L_y$ and the
respective boundary conditions as $BC_x$ and $BC_y$. We focus on the limit of
infinite length, $L_x \to \infty$, for which the results are independent of the
longitudinal boundary conditions.  For an infinite-length strip of a lattice
$\Lambda$, as the width $L_y \to \infty$, one expects $\langle k \rangle$ to
approach a limiting function of $p$ which is independent of the transverse
boundary conditions and is equal to $\langle k \rangle$ for the corresponding
infinite two-dimensional lattice $\Lambda$.  In particular, for a given
infinite-length, finite-width strip of the lattice $\Lambda$, it is of interest
to evaluate our exact expressions for $\langle k \rangle$ at $p=p_{c,\Lambda}$
and study how the resultant value approaches the critical value $\langle
k \rangle_{c,\Lambda}$ for the corresponding infinite two-dimensional lattice.

\section{Calculational Method}

For a given graph $G=(V,E)$ we calculate $\langle k \rangle$ by making use of
the fact that it can be expressed as a certain derivative of the (reduced) free
energy of the Potts model.  We first recall this expression.  Define a spanning
subgraph $G^\prime=(V,E^\prime)$ as a subgraph of $G$ with the same vertex set
$V$ and a subset of the edge (bond) set, $E^\prime \subseteq E$.  From the
definition of the average number of clusters per site, we have
\beqs
\langle k \rangle & = & \frac{(1/n)\sum_{G^\prime} k(G^\prime)
p^{e(G^\prime)}(1-p)^{e(G)-e(G^\prime)} }{
\sum_{G^\prime} p^{e(G^\prime)}(1-p)^{e(G)-e(G^\prime)} } \cr\cr
& & \cr\cr
& = &
\frac{(1/n)\sum_{G^\prime} k(G^\prime)[p/(1-p)]^{e(G^\prime)}}{
\sum_{G^\prime} [p/(1-p)]^{e(G^\prime)}} \ . 
\label{k}
\eeqs
This follows because each $G^\prime$ contains $k(G^\prime)$ connected
components, and appears in the numerator of the expression in the first line
with weight given by $p^{e(G^\prime)}(1-p)^{e(G)-e(G^\prime)}$, since the
probability that all of the bonds in $G^\prime$ are present is
$p^{e(G^\prime)}$ and the probability that all of the other $e(G)-e(G^\prime)$
bonds in $G$ are absent is $(1-p)^{e(G)-e(G^\prime)}$.  This sum in the
numerator over the set of spanning subgraphs $G^\prime$ is normalized by the
indicated denominator and by an overall factor of $1/n$ to obtain the average
number of connected components (clusters) per site.  As noted, we shall focus
on the limit in which the strip length $L_x \to \infty$ and hence 
$n \to \infty$.  

The cluster representation for the partition function of the $q$-state
Potts model is \cite{kf,wurev}
\beq
Z(G,q,v) = \sum_{G^\prime \subseteq G} q^{k(G^\prime)}v^{e(G^\prime)} \ . 
\label{cluster}
\eeq
In the Potts spin model, the spin $\sigma_i$ at each vertex $i$
can take on $q$ different values $\sigma_i=1,2,..,q$ and
\beq
v \equiv e^K - 1 \ , \quad K = \beta J \ , 
\label{v}
\eeq
where $\beta=1/(k_BT)$ with $T$ the temperature and $J$ the spin-spin
exchange constant.  The formula (\ref{cluster}) allows one to generalize $q$
 from the positive integers to a non-negative real variable while retaining a 
Gibbs measure for $Z(G,q,v)$. 

On a finite graph $G$ one defines the (reduced) free energy per site of the
Potts model as
\beq
f(G,q,v) = \ln [ Z(G,q,v)^{1/n} ] 
\label{fn}
\eeq
and, in the limit $n \to \infty$, 
\beq
f(\{G\},q,v) = \lim_{n \to \infty} f(G,q,v)
\label{f}
\eeq
where we use the symbol $\{G\}$ to denote the formal limit $\lim_{n \to
\infty}G$ for a given family of graphs.  (The actual free energy per site is 
$F = -k_BT f$; henceforth we shall simply refer to $f$ as the free energy.) 
If one sets
\beq
v = v_p \equiv \frac{p}{1-p} \ , 
\label{vp}
\eeq
differentiates $f(G,q,v_p)$ with respect to $q$, and sets $q=1$, one obtains
precisely the expression for $\langle k \rangle_n$, as given in eq. (\ref{k}),
i.e.,
\beq
\left . \langle k \rangle_n = \frac{\partial f(G,q,v_p)}{\partial q} 
\right |_{q=1} \ . 
\label{kdfdqn}
\eeq
In particular, in the limit $n \to \infty$, 
\beq
\left . 
\langle k \rangle = \frac{\partial f(\{G\},q,v_p)}{\partial q} 
\right |_{q=1} \ . 
\label{kdfdq}
\eeq
Our method for calculating $\langle k \rangle$ is to use eq. (\ref{kdfdq}) in
conjunction with exact results that we have computed for the free energy of the
Potts model on infinite-length, finite-width strips of various lattices
\cite{a}-\cite{zt}.  One can also study quantities in the percolation problem
in an analytic manner without making use of eq. (\ref{kdfdq}) (e.g.,
\cite{derrida1}), but the method we use suffices for our purposes.  One of the
interesting properties of the formula (\ref{kdfdq}) is that it relates a
geometric property of (the $n \to \infty$ limit of) a bond-diluted graph with
the derivative of the Potts model, evaluated at a certain temperature as $q \to
1$, on a graph with no bond dilution.  Specifically, the mapping (\ref{vp}), in
conjunction with eq. (\ref{kdfdq}), formally associates the interval $0 \le p
\le 1$ with the interval $0 \le v \le \infty$, which is the physical range of
values of the temperature variable $v$ for the ferromagnetic Potts model.  In
passing, we recall that the connection of percolation to the $q=1$ Potts model
implies that the percolation transition on 2D lattices is second-order, in the
same universality class as the latter model, with the exactly known thermal and
magnetic critical exponents $y_t=3/4$ and $y_h=91/48$ and thus
$\alpha=\alpha^\prime=-2/3$, $\beta=5/36$, $\gamma=\gamma^\prime=43/18$,
$\delta=91/5$, etc., the values of which were later illuminated further by the
application of conformal algebra.

\section{Some Basic Properties of $\langle \lowercase{k} \rangle$ }

As background for our exact results, it will be useful to note some basic
properties of $\langle k \rangle$.  These can all be understood using
elementary methods. First, in the limit as $p \to 0$, the only nonzero
contribution to the sum in the numerator, and to the sum in the denominator, of
eq. (\ref{k}) arises from the spanning subgraph $G^\prime$ with no edges,
$e(G^\prime)=\emptyset$, which has $n$ connected components, so $\lim_{p \to 0}
\langle k \rangle_n = \lim_{p \to 0} \langle k \rangle = 1$.  Second, in the
limit as $p \to 1$, the only nonzero contribution to the sum in the numerator,
and to the sum in the denominator, of eq. (\ref{k}) arises from the term in
which the spanning subgraph $G^\prime$ is $G$ itself, with $k(G)=1$, so
$\lim_{p \to 1} \langle k \rangle_n = 1/n$ and hence $\lim_{p \to 1} \langle k
\rangle = 0$.  Third, for a given $\{G\}$, the average per-site cluster number
$\langle k \rangle$ on $\{G\}$ is a monotonically decreasing function of $p$
for $p \in [0,1]$.  To see this, we start with $\langle k \rangle_n$ on a
finite $G$ and observe that increasing $p$ in the interval $[0,1]$ decreases
the number of components $k(G)$.  Dividing by $n$ and taking the limit $n \to
\infty$ yields the corresponding monotonicity result for $\langle k \rangle$ on
$\{ G \}$.

Further, for the infinite-length, finite-width strips considered here, one can
show that $\langle k \rangle$ is a (real) analytic function of $p$ for $p \in
[0,1)$ as follows.  We use the general relation eq. (\ref{kdfdq}) and observe
that the mapping (\ref{vp}) formally associates the value $p=0$ in the bond
percolation problem with the value $v=0$, i.e., infinite temperature, in the
Potts model free energy.  Since the free energy of a spin model is always
analytic near $\beta=0$, i.e., $v=0$, it follows that for any $\{G\}$, $\langle
k \rangle$ is analytic in the neighborhood of $p=0$.  As $p$ increases from 0
to 1, the value of $v$ corresponding to it via the mapping (\ref{vp}) increases
 from $v=0$ to $v=\infty$ (i.e., $T=0$).  Since infinite-length, finite-width
lattice strip graphs are quasi-one-dimensional, we can use the property that
the free energy of a quasi-one-dimensional spin system with finite-range
spin-spin interactions, such as the Potts model, is analytic for all finite
temperatures.  It follows that $\langle k \rangle$ is analytic for $0 \le p <
1$. In general, for quasi-1D systems, the critical percolation probability is
$p_c=1$, so that many quantities in 1D percolation are nonanalytic at $p=1$;
examples include the average cluster size function $S(p)=(1+p)/(1-p)$, which
diverges as $p \to 1^-$, and the percolation probability $P(p)$, which is zero
for $0 \le p < 1$ and jumps discontinuously to 1 at $p=1$
\cite{bethe,stanley77,stauffer}.  This is analogous to the fact that the
critical temperature for ferromagnetic spin models with short-range spin-spin
interactions in 1D is $T_c=0$ ($v=\infty$).  However, for all of the cases that
we have calculated, the mean cluster number $\langle k \rangle$, is
analytic at $p=1$ as well as for $p \in [0,1)$.  This is somewhat similar to
the situation with the internal energy of a ferromagnetic 1D spin system with
short-range spin-spin interactions, which is analytic at $T=0$ even though the
susceptibility and correlation lengthy diverge as $T \to 0$ and the
magnetization jumps from 0 to 1 at $T=0$.  This analyticity of $\langle
k \rangle$ for $p \in [0,1]$ for infinite-length, finite-width lattice strips
is, of course, different from the situation on lattices $\Lambda$ of
dimensionality $d \ge 2$, where $\langle k \rangle$ generically has a (finite)
nonanalyticity at $p=p_{c,\Lambda}$.  The fact that $\langle k \rangle$ for the
infinite-length, finite-width lattice strips is analytic for $p \in [0,1]$ does
not prevent this quantity from having singularities at unphysical 
values of $p$, including real values outside the interval $0 \le p \le 1$ and
complex values.  These will be discussed below.

The property that $\langle k \rangle$ (calculated in the limit as $L_x \to
\infty$) is independent of the longitudinal boundary conditions imposed on the
lattice strip follows from the same property for the Potts model free energy.
As noted above, we further expect that, for a given transverse boundary
condition and for a given $p \in [0,1]$, the value of $\langle k \rangle$ for
an infinite-length strip of the lattice $\Lambda$ approaches the corresponding
value of $\langle k \rangle$ for the (infinite) two-dimensional lattice
$\Lambda$ as $L_y \to \infty$.  This means that for moderate finite widths
$L_y$, the resultant $\langle k \rangle$ can serve as exactly calculated
approximation to the value of $\langle k \rangle$ for the corresponding
two-dimensional lattice.  Of course, this approximation refers to the value
only, not the full analytic structure; as we have shown, for any finite $L_y$,
regardless of how large, $\langle k \rangle$ does not contain the (finite)
nonanalyticity at $p_{c,\Lambda}$ that is present in $\langle k \rangle$ for
the infinite two-dimensional lattice.

\section{Relevant Results for Bond Percolation on Two-Dimensional Lattices}

\subsection{Values of $p_{c,\Lambda}$} 

Here we briefly recall some results that we shall use for bond percolation on
two-dimensional lattices.  For a planar lattice $G$, the (planar) dual, $G^*$
is the lattice whose vertices and faces are given, respectively, by the faces
and vertices of $G$ and whose edges link the faces of $G$.  For the
thermodynamic limit of a regular two-dimensional lattice graph $G$, the
critical probability $p_c$ for bond percolation satisfies \cite{exactpc}
$p_{c,\{G\}} + p_{c,\{G^*\}}=1$.  In particular, it follows that, since the
square (sq) lattice is self-dual, $p_{c,sq} = 1/2$.  The values of $p_c$ are
also known exactly for the triangular (tri) and honeycomb (hc) lattices
\cite{exactpc}: $p_{c,tri} = 2\sin(\pi/18) = 0.347296..$ and $p_{c,hc} =
1-p_{c,tri} = 0.6527036..$.

It is useful to recall that for a general lattice $\Lambda$, $p_{c,\Lambda}$
can be obtained from a knowledge of the critical temperature of the $q \to 1$
Potts ferromagnet.  The relation (\ref{vp}) is equivalently written as
$p=v/(1+v)$.  Hence, the critical percolation probability is determined as
$p_{c,\Lambda} = v_{c,\Lambda}/(1+v_{c,\Lambda})$, where $v_{c,\Lambda}$ is the
value of $v$ corresponding to the phase transition temperature of the Potts
ferromagnet on $\Lambda$ in the limit $q \to 1$.  The value of $v_{c,\Lambda}$
is known exactly for the square, triangular and honeycomb lattices
\cite{wurev}.  Thus, $v_c$ for the square lattice is given by the self-duality
relation $v=\sqrt{q}$; setting $q=1$ yields $v_{c,sq}=1$ and substituting this
into eq. (\ref{vp}) yields $p_{c,sq}=1/2$.  The value of $v_{c,tri}$ for the
triangular lattice is determined as the (unique) real positive solution of the
equation \cite{kimjoseph} $v^3+3v^2-q=0$. Solving this for $q=1$ and
subsituting into eq. (\ref{vp}) yields the expression above for $p_{c,tri}$.
 From the duality of the triangular and honeycomb lattices, one obtains the
critical equation for the latter from the former by the inversion map
$v/\sqrt{q} \to \sqrt{q}/v$, which yields $v^3-3qv-q^2=0$. Again setting $q=1$,
calculating the unique positive root of this equation, and substituting into
(\ref{vp}) gives the expression for $p_{c,hc}$. 

We next comment on $p_c$ for the kagom\'e lattice and first recall the
definition of this lattice. An Archimedean lattice is a uniform tiling of the
plane by one or more regular polygons with the property that every vertex is
equivalent to every other vertex.  An Archimedean lattice is specified by the
ordered product $\prod p_i^{a_i}$, where, as one makes a small circuit (say
clockwise) around any vertex, one traverses the sequence of polygons $p_i$,
with the possibility that a given $p_i$ occurs $a_i$ times in a row (for
notation, see, e.g., \cite{wn}). In this notation, the kagom\'e is defined as
$(3 \cdot 6 \cdot 3 \cdot 6)$.  Although the critical bond percolation
probability $p_{c,kag}$ is not known exactly, it has been determined
numerically to high precision \cite{ziffsuding97} as $p_{c,kag} =
0.5244053(3)$, where the number in parentheses is the estimated uncertainty in
the last digit.

\subsection{Values of $\langle k \rangle_{c,\Lambda}$} 

Because $\langle k \rangle$ has only a finite nonanalyticity at
$p=p_{c,\Lambda}$, one can obtain rough values of $\langle k
\rangle_{c,\Lambda}$ from Taylor series expansions around $p=0$ and $p=1$. 
Using results from \cite{templieb,baxtertemp}, Ref. \cite{ziff97} obtained 
explicit exact values of $\langle k \rangle_c$ for bond percolation on three
two-dimensional lattices: 
\beq
\langle k \rangle_{c,sq} = \frac{3^{3/2}-5}{2} = 0.0980762...
\label{kcsq}
\eeq
and
\beq
\langle k \rangle_{c,tri} = \frac{35}{4} - \frac{3}{p_{c,tri}} = 0.111844...
\label{kctri}
\eeq
and, using also \cite{es66}, 
\beq
\langle k \rangle_{c,hc} = \frac{1}{2}(\langle k \rangle_{c,tri} + p_{c,tri}^3) 
= 0.07686667...
\label{kchc}
\eeq
(Here we take into account that $\langle k \rangle$ is defined per site, while
the quantity $n_c^{B-HC}$ in \cite{ziff97} is defined per unit cell and there
are two sites per unit cell on the honeycomb lattice.)  Since 
$p_{c,kag}$ is not known exactly, neither is $\langle k \rangle_{c,kag}$.
We are not aware of numerical simulations of bond percolation on kagom\'e
lattices that have yielded an approximate value of $\langle k \rangle_{c,kag}$,
or of reasonably long series expansions that could be used to obtain an accurate
estimate of this quantity. 

\subsection{Series Expansions}

Taylor series expansions played an important role in the early investigation of
critical exponents and tests of scaling for percolation. These expansions were
performed about the points $p=0$ (low-density) and $p \to 1$ (high-density), or
equivalently, $r \to 0$, where
\beq
r= 1-p \ . 
\label{r}
\eeq
These series expansions have typically been calculated for one of two related
quantities, the average number, per site, of bond clusters, $\langle k_{bond}
\rangle$, or the average number, per bond ($pb$), of bond clusters, $\langle
k_{bond} \rangle_{pb}$.  For an (infinite) lattice $\Lambda$ with coordination
number $\kappa_\Lambda$, there are $\kappa_\Lambda/2$ bonds per site, so
$\langle k_{bond} \rangle = (\kappa_\Lambda/2) \langle k_{bond} \rangle_{pb}$.
The average number of components (= site clusters) per site, $\langle k
\rangle$, differs from the average number of bond clusters per site, $\langle
k_{bond} \rangle$ in that $\langle k \rangle$ counts all components, including
isolated sites, whereas $\langle k_{bond} \rangle$ counts the bond clusters
that have a nonzero number of bonds and therefore excludes isolated
sites. Hence, for an (infinite) lattice $\Lambda$,
\beq 
\langle k \rangle_\Lambda = \langle k_{bond} \rangle_\Lambda +
(1-p)^{\kappa_\Lambda} \ . 
\label{sitebondrel}
\eeq
Clearly, $\langle k_{bond} \rangle \to 0$ rather than 1 as $p \to 0$, since no
bond clusters with a nonzero number of bonds are present on $G$ in this limit.

Early calculations of series for the average bond cluster number $\langle
k_{bond} \rangle$ were carried out in Refs. \cite{exactpc,es66} and, to higher
order in \cite{sykesseries} for the square, triangular, and honeycomb
lattices. These have the form
\beq
\langle k_{bond} \rangle = \sum_{s=1}^\infty p^s D_s(r)
\label{kbseries}
\eeq
where $r$ is given by eq. (\ref{r}) and the $D_s$ are the perimeter
polynomials.  Using the relation (\ref{sitebondrel}), it is 
straighforward to calculate series for $\langle k \rangle$ from those for 
$\langle k_{bond} \rangle$.  For our present purposes, we shall need only the
first few terms of these series. For $p \to 0$ these are 
\beq
\langle k \rangle_{sq} = 1-2p+p^4+2p^6-2p^7+7p^8+O(p^9)
\label{sqseriesp0}
\eeq
\beq
\langle k \rangle_{tri} = 1-3p+2p^3+3p^4+3p^5+3p^6+6p^7+O(p^9) 
\label{triseriesp0}
\eeq
\beq
\langle k \rangle_{hc} = 1 - \frac{3}{2}p + \frac{1}{2}p^6 + 
\frac{3}{2}p^{10} +O(p^{11}) \ . 
\label{hcseriesp0}
\eeq
Thus, for small $p$, $\langle k \rangle$ is linear, with slope 
$-\kappa_\Lambda/2$, so that $\langle k \rangle$ decreases more rapidly on a
lattice with larger coordination number. 

For the $p \to 1$ series, we have 
\beq
\langle k \rangle_{sq} = r^4+2r^6-2r^7+7r^8+O(r^9)
\label{sqseriesp1}
\eeq
\beq
\langle k \rangle_{tri} = r^6+3r^{10}-3r^{11}+2r^{12}+O(r^{14})
\label{triseriesp1}
\eeq 
\beq
\langle k \rangle_{hc} = r^3 + \frac{3}{2}r^4 + \frac{3}{2}r^5 + 
\frac{3}{2}r^6 + O(r^7) \ . 
\label{hcseriesp1}
\eeq
These have the general form $\langle k \rangle_\Lambda = r^{\kappa_\Lambda} +
...$ as $p \to 1$, so that the larger the coordination number is, the smaller
$\langle k \rangle$ in these expansions.

Because of the significant finite-width effects on the lattice strips
considered here, one does not expect the Taylor series expansions of the exact
expressions for $\langle k \rangle$ on these strips to match many orders of the
expansions for the corresponding two-dimensional lattices.  However, our exact
results can give new insight into one property of these series.  In early work
it was found that the radii of convergence of Taylor series expansions around
both $p=0$ and $p=1$ were typically set by unphysical singularities, and these
radii of convergence were less than the distance from the expansion point to
the physical singularity, $p_{c,\Lambda}$, for the small-$p$ expansions and
$r_{c\Lambda} = 1-p_{c, \Lambda}$ for small-$r$ expansions \cite{sykesseries}.
(This was also the case for three-dimensional lattices \cite{sykes3D}.)  This
is reminiscent of the situation for low-temperature Taylor series expansions
for various discrete spin models such as the Ising model, whose radii of
convergence in the expansion variable $z = e^{-\beta J}$ were often smaller
than the distance from the origin to the actual critical point $z_c$, again as
a consequence of complex-temperature singularities \cite{lowtemp,ms}.  Although
spin models with finite-range spin-spin interactions only have possible phase
transitions at $T=0$, the complex-temperature singularities of the free energy
for such quasi-1D systems does exhibit some similarities with those of 2D
systems \cite{ms}-\cite{potts}.  These complex-temperature singularities are
associated with complex-temperature phase boundaries ${\cal B}$, which are the
continuous accumulation set of the (Fisher \cite{fisher}) zeros of the
partition function.  In the same way, we can use our exact calculations of
$\langle k \rangle$ for infinite-length, finite-width lattice strips to gain
some insight into the unphysical singularities encountered in the
above-mentioned series studies for percolation.

\section{Strips of the Square Lattice}

\subsection{$L_y=1$}
                   
The well-known result
\beq
\langle k \rangle_{1D}=1-p
\label{k1d}
\eeq
for the infinite line can be derived directly using probability methods.  Here
we illustrate how it can be derived via eq. (\ref{kdfdq}).  An elementary
calculation yields the Potts free energy $f(1D,q,v)=\ln(q+v)$.  Using
eq. (\ref{kdfdq}) yields the above result for $\langle k \rangle_{1D}$.  This
has the value 1/2 at $p=p_{c,sq}$ (see Table \ref{kcvalues}).  Note that the
bond cluster number per site in 1D is $\langle k_{bond} \rangle = p(1-p)$, in
accord with the relation (\ref{sitebondrel}).

\subsection{Free Transverse Boundary Conditions}

For $L_y \ge 2$, we label an infinite-length strip of width $L_y$ of the
lattice $\Lambda$ with given transverse boundary conditions $BC_y$ as
$\Lambda,(L_y)_{BC_y}$.  In particular, the $L_y=2$ square-lattice strips with
free (F) and periodic (P) transverse boundary conditions are denoted $sq,2_F$
and $sq,2_P$.

\subsubsection{$2_F$}

The free energy of the Potts model for the $sq,2_F$ strip is \cite{a}
\beq
f(sq,2_F,q,v) = \frac{1}{2}\ln \lambda_{sq,2F,1}
\label{fsqxy2}
\eeq
where 
\beq
\lambda_{sq,2F,j} =\frac{1}{2} \Bigl ( T_{s2F} \pm \sqrt{R_{s2F}} \ \Bigr )
\label{lambda_sq2F}
\eeq
with $j=1,2$ corresponding to $\pm$ and 
\beq
T_{s2F}=v^3+4v^2+3qv+q^2
\label{ts12}
\eeq
\beqs
R_{s2F} & = & v^6+4v^5-2qv^4-2q^2v^3+12v^4 \cr\cr
        & + & 16qv^3+13q^2v^2+6q^3v+q^4 \ . 
\label{rs12}
\eeqs
In eq. (\ref{lambda_sq2F}) only the $j=1$ term is relevant for the free
energy, while the $j=2$ term will be discussed below.

 From eq. (\ref{fsqxy2}) we calculate the average cluster number per site
\beq
\langle k \rangle_{sq,2_F} = \frac{(1-p)^2(2+p-2p^2)}{2(1-p^2+p^3)} \ . 
\label{ksqxy2}
\eeq
This is plotted in Fig. \ref{sq} together with cluster numbers calculated for
other strips.  At $p=p_{c,sq}=1/2$, this average cluster number has the value
$\langle k \rangle_{sq,2_F} = 2/7 \simeq 0.28571$. 

 From the exact expression for $\langle k \rangle_{sq,2_F}$ we compute 
the respective Taylor series expansions 
\beq
\langle k \rangle_{sq,2_F} = 1-\frac{3}{2}p + \frac{1}{2}p^4 + \frac{1}{2}p^6 +
O(p^7) \quad \text{for} \quad p \to 0
\label{ksqxy2_taylorp0}
\eeq
and, in terms of the variable $r$ in eq. (\ref{r}), 
\beq
\langle k \rangle_{sq,2_F} = \frac{1}{2}r^2 + 2r^3 -\frac{7}{2}r^5 + O(r^6)
 \quad \text{for} \quad r \to 0 \ . 
\label{ksqxy2_taylorp1}
\eeq
Thus for this strip $\langle k \rangle$ is linear for small $p$ and
vanishes quadratically as $p \to 1$.  As expected for such a small width, these
series differ from the series for the square lattice, although the linear
behavior for small $p$ is common to both. 

The expression for $\langle k \rangle$ for this strip has singularities, which
are simple poles, where the denominator $1-p^2+p^3=0$, at 
\beq
p \simeq -0.7549 \ , \quad 0.8774 \pm 0.7449i \ .
\label{sq2Fpoles}
\eeq
The first of these poles is the closest to the origin and determines the radius
of convergence of the small-$p$ Taylor series in eq. (\ref{ksqxy2_taylorp0}) to
be approximately 0.7549.  The complex pair are the same distance from the point
$p=1$ and imply that this series converges for $|1-p| \lsim 0.7549$.  Thus,
although $\langle k \rangle_{sq,2_F}$ is an analytic function of $p$ for $p \in
[0,1]$, the Taylor series expansions about $p=0$ and $p=1$ have radii of
convergence less than unity because of singularities of this function at 
real and complex values outside the physical interval $0 \le p \le 1$.  It is
interesting that although $\lambda_{sq,2_F}$ is an algebraic function of $v$
(hence $p$, via eq. (\ref{vp}), the resultant expression for $\langle k
\rangle_{sq,2_F}$ is a rational function of $p$.  However, this is a
consequence of the small value of $L_y$.  The same comment applies to the
property that $\langle k \rangle_{sq,2_F}$ is meromorphic, i.e., its only
singularities are simple poles.  As will be seen, these features are also true
of the cluster number $\langle k \rangle$ for other $L_y=2$ strips considered
here.

In this very simple context of a quasi-1D strip, one hence gains some insight
into the similar influence of unphysical singularities in series expansions
about $p=0$ and $p=1$ for percolation on higher-dimensional lattices.  To
understand these poles more deeply, we observe that although the free energy
$f(sq,2_F,q,v)$ only depends on $\lambda_{sq,2F,1}$, the partition function for
free longitudinal boundary conditions \cite{sqpxy} and $q \ne 1$ in general is
a symmetric sum of $L_x$'th powers of both of the $\lambda_{sq,2F,j}$'s for
$j=1$ and $j=2$ (given as eq. (5.17) of Ref. \cite{a}).  We are interested in
the limit $q \to 1$.  It is necessary to take account of a subtlety concerning
the dependence of the complex-$v$ phase boundary ${\cal B}$ of the Potts model,
as a function of $q$.  In previous work (see eqs. (2.8)-(2.12) of
Ref. \cite{a}, eq. (1.9) of Ref. \cite{w})) we have pointed out the
noncommutativity at certain special values of $q_s$, including $q_s=0,1$,
namely
\beq
\lim_{n \to \infty} \lim_{q \to q_s} Z(G,q,v)^{1/n} \ne
\lim_{q \to q_s} \lim_{n \to \infty} Z(G,q,v)^{1/n} 
\label{fnoncomm}
\eeq
and we have noted that because of this noncommutativity, for 
the special set of points $q=q_s$ one must distinguish between (i) $({\cal
B}(\{G\},q_s))_{nq}$, the continuous accumulation set of the zeros of
$Z(G,q,v)$ obtained by first setting $q=q_s$ and then taking $n \to \infty$,
and (ii) $({\cal B}(\{G\},q_s))_{qn}$, the continuous accumulation set of the
zeros of $Z(G,q,v)$ obtained by first taking $n \to \infty$, and then taking $q
\to q_s$.  For these special points,
\beq
({\cal B}(\{G\},q_s))_{nq} \ne ({\cal B}(\{G\},q_s))_{qn} \ .
\label{bnoncomm}
\eeq
A previous case of this was the $q=2$ (Ising) special case of the Potts model.
Indeed, in that case it was noted that ${\cal B}_{qn}$ does not have the
inversion symmetry $e^K \to e^{-K}$ that characterizes the Ising model and its
complex-temperature phase boundary ${\cal B}_{nq}$ for a bipartite lattice (see
pp. 396, 433-435 of Ref. \cite{a}).  This noncommutativity is also present at
the value $q=1$ relevant for percolation.  If one uses the definition ${\cal
B}_{nq}$ with $q=1$ for percolation, as one uses the definition ${\cal B}_{nq}$
with $q=2$ for the Ising model, then while ${\cal B}_{nq}$ is nontrivial for
the Ising model, ${\cal B}_{nq}$ is trivial for the percolation problem.  The
reason for this is that if one sets $q=1$ first, then, from the Hamiltonian
definition of the Potts model, since the spins are the same on all sites, the
spin-spin interactions on each bond contribute a factor $e^K$ to the partition
function, so one has the elementary result
\beq 
Z(G,1,v)=e^{K \, e(G)} = (v+1)^{e(G)}
\label{zq1}
\eeq
Substituting $v=v_p$ as in eq. (\ref{vp}) gives
\beq
Z(G,1,v_p)= (1-p)^{-e(G)}
\label{zg1vp}
\eeq
Evidently, $Z(G,1,v_p)$ has no zeros, so that ${\cal B}_{nq}=\emptyset$ in the
complex $p$ plane.  Equivalently, $Z(G,1,v)$ has only a single zero at the 
point $v=-1$, which maps, via eq. (\ref{vp}), to the circle at infinity
in the complex $p$ plane.  We have noted above that in general, for $q \ne 1$,
the partition function of the Potts model consists of a symmetric sum of
$L_x$'th powers of $\lambda_{sq,2_F,1}$ and $\lambda_{sq,2_F,2}$; if one sets
$q=1$, the coefficient of $(\lambda_{sq,2_F,2})^{L_x}$ vanishes, and the
$(\lambda_{sq,2_F,1})^{L_x}$ term, with its coefficient, reduces to the form
(\ref{zg1vp}) (where in the labelling convention of Ref. \cite{a}, $L_x+1$
denotes the number of squares on the $sq,2_F$ strip). 

However, the fact that eq. (\ref{kdfdq}) involves a derivative means that it is
sensitive to properties of the Potts model in the neighborhood of the point
$p=1$ as well as at this point.  This suggests that one consider the possible
role of the locus ${\cal B}_{qn}$, although one must use caution in doing this
because of the noncommutativity discussed above. Below we shall use the
notation ${\cal B}_{qn}$ to mean specifically the boundary defined for
$n\to\infty$ and $q\to 1$, relevant to the percolation problem.

We find some intriguing connections between the locus ${\cal B}_{qn}$ and
complex-$p$ singularities in $\langle k \rangle$.  Let us first calculate
${\cal B}_{qn}$ for the $sq,2_F$ strip as $q \to 1$ but with $q \ne
1$. Evaluating eq. (\ref{lambda_sq2F}) for $q \to 1$, we obtain
\beq
\lambda_{sq,2F,1} = \frac{1}{(1-p)^3} 
\label{lam_sq2F1}
\eeq
and
\beq
\lambda_{sq,2F,2} = \frac{p^2}{(1-p)^2} \ . 
\label{lam_sq2F2}
\eeq
The locus ${\cal B}_{qn}$ is the set of solutions of the
equation of degeneracy in magnitude of dominant $\lambda$'s.  This locus can
be seen as a special case of the more general phase boundary for the Potts
model in the $v$ plane for a fixed $q$, or in the $q$ plane for fixed $v$. For
the present case, since there are only two $\lambda_{sq,2F,j}$'s, for $j=1,2$,
this equation is $|\lambda_{sq,2F,1}|=|\lambda_{sq,2F,2}|$, i.e.,
\beq
|p^2(1-p)|=1 \ . 
\label{sq2Feq}
\eeq
In terms of the polar coordinates $p=\rho e^{i\theta}$ this equation reads
$\rho^4(1+\rho^2-2\rho \cos \theta)=1$.  The solution is a closed egg-shaped
curve, shown in Fig. \ref{sq2fb}, that crosses the real $p$ axis at $p\simeq
-0.7549$ and $p \simeq 1.466$ and the imaginary $p$ axis at $p \simeq \pm
0.8688i$.  This thus constitutes the phase boundary in the complex $p$ plane,
separating this plane into two regions.  As follows from the general discussion
above, the physical interval $0 \le p \le 1$ lies entirely in one phase.  The
three poles of $\langle k \rangle_{sq,2_F}$ listed in eq. (\ref{sq2Fpoles}) lie
on this boundary ${\cal B}_{qn}$.

\subsubsection{$3_F,4_F,5_F$}

\begin{figure}[hbtp]
\centering
\leavevmode
\epsfxsize=2.2in
\begin{center}
\leavevmode
\epsffile{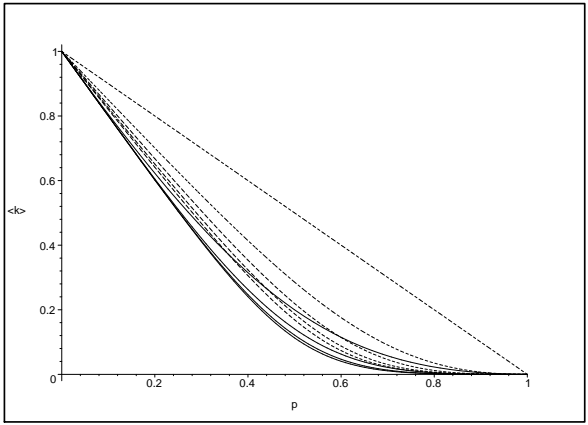}
\end{center}
\vspace{-10mm}
\caption{\footnotesize{Plots of $\langle k \rangle$ (vertical axis) as a
function of $p \in [0,1]$ (horizontal axis) for infinite-length, finite-width
strips of the square lattice.  The dashed and solid curves refer to free and
periodic transverse boundary conditions, respectively. For a given $p$, the
dashed curves are, in order of descending value of $\langle k \rangle$, for
$1_F \le (L_y)_F \le 5_F$, and the solid curves are, in the same order, for
$2_P \le (L_y)_P \le 5_P$.}}
\label{sq}
\end{figure}

\begin{figure}[hbtp]
\centering
\leavevmode
\epsfxsize=2.2in
\begin{center}
\leavevmode
\epsffile{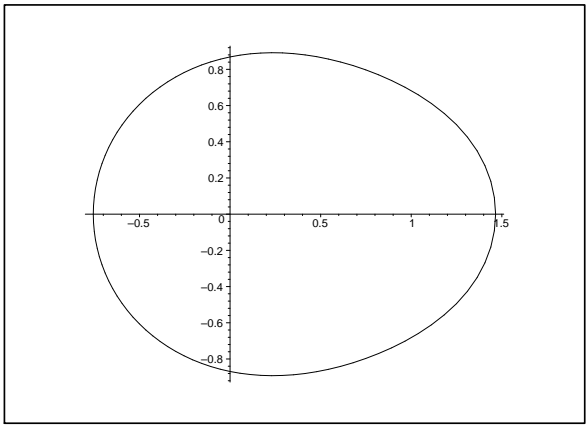}
\end{center}
\vspace{-10mm}
\caption{\footnotesize{Plot of the boundary ${\cal B}_{qn}$ in the complex $p$
plane for the infinite-length $sq,2_F$ lattice strip. Horizontal and vertical 
axes are $Re(p)$ and $Im(p)$.}}
\label{sq2fb}
\end{figure}

The Potts model free energy $f$ for the infinite-length $sq,3_F$ strip was
calculated in Ref. \cite{s3a}.  The free energy is given by
$f(sq,3_F,q,v)=(1/3)\ln \lambda_{sq,3_F}$, where $\lambda_{sq,3_F}$ is the
(maximal) root of an algebraic equation of degree 4.  Because of the
complicated nature of the expression for this quartic root, we do not present
it here. We have calculated $f(sq,3_F,q,v)$, and hence $\langle k
\rangle_{sq,3_F}$, to high precision by numerically solving for $f(sq,3_F,q,v)$
for a range of values of $q$ near unity, for each value of $p$, and carrying
out the differentiation in eq. (\ref{kdfdq}).  Although this is numerical, the
computational steps can be carried out with almost arbitrarily high precision,
so that, in practice, it is essentially equivalent to evaluating an explicit
exact analytic expression.  We also apply this procedure for larger strip
widths, using the exact calculation of $f$ for $sq,4_F$ and $sq,5_F$ in
Ref. \cite{ts} (see also \cite{zt}).  The resulting values of $\langle k
\rangle$ are plotted as functions of $p$ in Fig. \ref{sq}, and the values of
$\langle k \rangle$ at $p=p_{c,sq}=1/2$ are listed in Table \ref{kcvalues}.
One could carry out similar calculations of $\langle k \rangle$ for larger
values of $L_y$, but our results are sufficient to show the nature of the
approach of $\langle k \rangle$ on these infinite-length, finite-width strips
to the average cluster number for the corresponding infinite two-dimensional
lattice.  Indeed, one of the most interesting pieces of information that we get
from our results - the exact determination of singularities of $\langle k
\rangle$ in the complex $p$ plane and their effect on the radii of convergence
of series expansions, can only be obtained for strip widths that are small
enough so that we can get exact explicit analytical forms for $\langle k
\rangle$.

\subsection{Periodic Transverse Boundary Conditions}
                                                                                
\subsubsection{$2_P$}

By using periodic transverse boundary conditions, one minimizes finite-width
effects in this transverse direction.  We consider first the $sq,2_P$ strip.
Note that this strip has double transverse bonds.  The free energy was computed
in Ref. \cite{s3a} and is given by
\beq
f(sq,2_P,q,v)=\frac{1}{2} \ln \lambda_{sq,2P,1}
\label{fsqxpy2}
\eeq
where
\beq
\lambda_{sq,2_P,j} = \frac{1}{2}\Bigl ( T_{s2P} \pm \sqrt{R_{s2P}} \ \Bigr )
\label{lambda_sq2p}
\eeq
with $j=1,2$ corresponding to $\pm$ and 
\beq
T_{s2P}=6v^2+4qv+q^2+4v^3+qv^2+v^4
\label{tsqcyl}
\eeq
and
\beqs
R_{s2P} & = & (v^4+6v^3+8v^2+3qv^2+6qv+q^2) \cr\cr
& \times & (v^4+2v^3+4v^2-qv^2+2qv+q^2) \ . 
\label{rsqcyl}
\eeqs
 From this, using eq. (\ref{kdfdq}), we calculate
\beq
\langle k \rangle_{sq,2_P} = \frac{(1-p)^2(2-3p^2+2p^3)}{2(1+p-p^2)(1-p+p^2)} 
\ . 
\label{ksqxpy2}
\eeq
At the value of $p_{c,sq}=1/2$ for the infinite square lattice this has the
value $\langle k \rangle_{sq,2_P} =1/5$. 

The expression (\ref{ksqxpy2}) has poles where $1+p-p^2$ vanishes, at
\beq
p_{1,2} = \frac{1}{2}(1 \pm \sqrt{5} \ ) \simeq 1.618 \ , \quad -0.6180
\label{sq2Ppolea}
\eeq
and where $1-p+p^2$ vanishes, at
\beq
p_{3,4}=\frac{1}{2}(1 \pm \sqrt{3} \ i) \simeq 0.5 \pm 0.866i \ .
\label{sq2Ppoleb}
\eeq
The second and first of these poles are closest to the points $p=0$ and $p=1$
and determine the radii of convergence of the respective Taylor series
expansions about these points both to be 0.618.  These expansions are
\beq
\langle k \rangle_{sq,2_P} = 1 - 2p + \frac{1}{2}p^2 + 2p^4 +O(p^5)
\label{ksqxpy2_taylorp0}
\eeq
and
\beq
\langle k \rangle_{sq,2_P} = \frac{1}{2}r^2 + 2r^4 - 2r^5 + O(r^6) \ . 
\label{ksqxpy2_taylorp1}
\eeq
Note that, in accord with the fact that the coordination number of this and any
infinite-length lattice strip of the square lattice with periodic transverse
boundary conditions is 4, the coefficient of the linear term in the small-$p$
expansion is equal to that of the expansion for the full square lattice. 

\begin{figure}[hbtp]
\centering
\leavevmode
\epsfxsize=2.2in
\begin{center}
\leavevmode
\epsffile{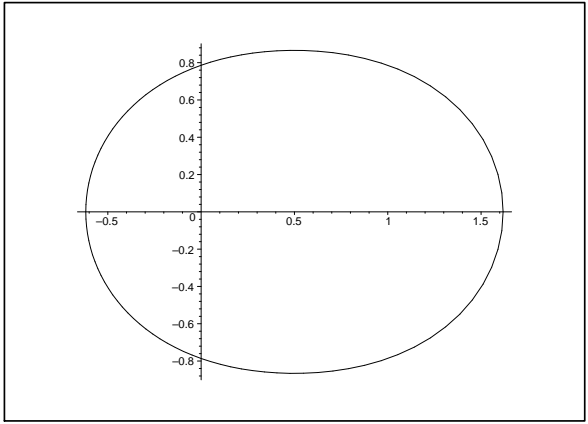}
\end{center}
\vspace{-10mm}
\caption{\footnotesize{Plot of the boundary ${\cal B}_{qn}$ in the complex $p$
plane for the infinite-length $sq,2_P$ lattice strip. Horizontal and vertical
axes are $Re(p)$ and $Im(p)$.}}
\label{sq2pb}
\end{figure}

We next discuss the connection of the poles in $\langle k \rangle_{sq,2_P}$
with the locus ${\cal B}_{qn}$.  Although $f(sq,2_P,q,v)$ only depends on the
quantity $\lambda_{sq,2P,1}$, the Potts model partition function for $ q \ne 1$
involves a symmetric sum of $L_x$'th powers of both $\lambda_{sq,2P,1}$ and
$\lambda_{sq,2P,2}$ \cite{ts}.  Evaluating $\lambda_{sq,2P,j}$ for $q \to 1$,
we have
\beq
\lambda_{sq,2P,1} = \frac{1}{(1-p)^4}
\label{lamsq2p1}
\eeq
\beq
\lambda_{sq,2P,2} = \frac{p^2}{(1-p)^2}
\label{lamsq2p2}
\eeq
The locus ${\cal B}_{qn}$ is the set of solutions of the equation
$|\lambda_{sq,2P,1}|=|\lambda_{sq,2P,2}|$, i.e.
\beq
|p(1-p)|=1 \ . 
\label{sq2pbeq}
\eeq
In terms of the polar coordinates defined above, this equation reads
$\rho^2(1+\rho^2-2\rho \cos\theta)=1$.  The solution forms a closed oval curve
in the complex $p$, shown in Fig. \ref{sq2pb}, that crosses the real axis at
the points $p_{1,2}$ in eq. (\ref{sq2Ppolea}) and the imaginary axis at the
points $p \simeq \pm 0.78615i$.  As in the case of the $sq,2_F$ strip, this
curve separates the $p$ plane into two regions.  All of the four poles of
$\langle k \rangle_{sq,2_P}$ given in eqs. (\ref{sq2Ppolea}) and
(\ref{sq2Ppoleb}) lie on this curve ${\cal B}_{qn}$. This property - that the
singularities of $\langle k \rangle$ lie on ${\cal B}_{qn}$ - is analogous to
the property that the singularities of thermodynamic functions of spin models
lie on the complex-temperature phase boundaries for these models, as we have
studied in earlier work \cite{ms}-\cite{potts}.  Having pointed out the
connection between these singularities and the locus ${\cal B}_{qn}$, we shall,
for the strips considered below, just summarize the singularities of $\langle k
\rangle$.

\subsubsection{$3_P,4_P,5_P$}

For the $sq,3_P$ strip, $f(sq,3_P,q,v)=(1/3)\ln \lambda_{sq,3_P}$, where
$\lambda_{sq,3_P}$ is the (maximal) root of a cubic equation. 
Although it is possible to display an analytic result for $\langle k
\rangle_{sq,3_P}$, it is sufficiently cumbersome that we do not give it here.
It is an algebraic, rather than rational, function of $p$. 
We do display the small-$p$ expansion, which is
\beq
\langle k \rangle_{sq,3P} = 1 - 2p + \frac{1}{3}p^3 + p^4 + O(p^5) \ . 
\label{ksqxpy3taylorp0}
\eeq
The free energy $f$ was calculated for the $sq,4_P$ and $sq,5_P$ strips in
Ref. \cite{ts}, and $\lambda_{sq,4_P}$ and $\lambda_{5_P}$ are roots of
equations of too high a degree to allow an explicit analytic solution.
Accordingly, we compute $\langle k \rangle$ by the numerical procedure
discussed above.  Results are given in Fig. \ref{sq} and Table \ref{kcvalues}.

\subsection{Self-Dual Strips of the Square Lattice}
                                                                                
It is of interest to calculate $\langle k \rangle$ for strips of the square
lattice that maintain a property of the infinite square lattice, namely
self-duality.  The strips with free and periodic transverse boundary conditions
considered above are not self-dual.  However, one can construct a cyclic strip
that is self-dual by adding a single external site to a cyclic square-lattice
strip of width $L_y$ and then adding bonds connecting all of the sites on one
side of the strip to this single external site.  We denote a self-dual ($sd$)
strip of this type as $sq,(L_y)_{sd}$. Before presenting our calculations, a
remark is in order concerning $p_c$ for these strips.  The physical meaning of
$p_c$ for a usual infinite lattice is, as mentioned before, that for $p \ge
p_c$ there exists an percolation cluster linking two points that are an
arbitrarily large distance apart.  Now consider the simplest of the cyclic
self-dual lattice graphs, with $L_y=1$; this is a wheel graph, having a rim
forming a circuit and a central site ($ \sim$ axle) connected to the sites on
the rim by $L_x$ bonds forming spokes.  Evidently, even in the limit $L_x \to
\infty$, the maximum distance between any two sites on this lattice graph is 2
bonds; to get from any site on the rim to any other site, one takes a
minimum-distance route that goes inward along one spoke to the central site and
out again on another spoke to the other site.  Similarly, for any finite $L_y$,
even as $L_x \to \infty$ there is a maximal finite distance $2L_y$ bonds
between any two sites.  Therefore, although this family of cyclic lattice
strips does maintain the property of self-duality of the infinite square
lattice, the notion of a critical $p_c$ beyond which there is a percolation
cluster linking two sites arbitrarily apart is not applicable to it since no
sites are arbitrarily far apart.

\subsubsection{$1_{sd}$}

The free energy is \cite{jz,sdg}
\beq
f(sq,1_{sd},q,v)=\ln \lambda_{sd1}
\label{fsdg}
\eeq
where
\beq
\lambda_{sd1,j}=\frac{1}{2}( T_{sd1} + \sqrt{R_{sd1}} \ )
\label{lamsd1}
\eeq
with 
\beq
T_{sd1} = 3v+q+v^2
\label{tsd1}
\eeq
\beq
R_{sd1} = 5v^2+2vq+2v^3+q^2-2v^2q+v^4 \ . 
\label{rsd1}
\eeq
 From this we calculate
\beq
\langle k \rangle_{sq,1_{sd}} = \frac{(1-p)^3}{1-p+p^2} \ . 
\label{ksdg}
\eeq
We have $\langle k \rangle_{sq,1_{sd}} = 1/6$ at $p=p_{c,sq}$. 
The mean cluster number $\langle k \rangle$ in eq. (\ref{ksdg}) has the
following Taylor series expansions for $p \to 0$ and $p \to 1$: 
\beq
\langle k \rangle_{sq,1_{sd}}=1 - 2p + p^3 + p^4 - p^6 - p^7 + p^9 + O(p^{10}) 
\label{ksdgtaylorp0}
\eeq
\beq
\langle k \rangle_{sq,1_{sd}} = r^3 + r^4 - r^6- r^7 + r^9 + O(r^{10})  \ . 
\label{ksdgtaylorp1}
\eeq
One sees that the coefficient of the linear term in the small-$p$ expansion
correctly matches that of the series for the infinite square lattice and the
power of the leading-order term in $p\to 1$ expansion is 3, which, although not
equal to the power 4 in the corresponding expansion in eq. (\ref{sqseriesp1}),
is at least closer than the power of 2 for the $L_y=2$ square-lattice strips
with free or periodic boundary conditions.  The poles in eq. (\ref{ksdg}) at
$p=(1/2)(1 \pm \sqrt{3}i)$ set the radii of convergence of the small-$p$ and
small-$r$ expansions as unity in both cases, i.e., the full physical interval
$0 \le p \le 1$. 

\subsubsection{$2_{sd},3_{sd},4_{sd}$}

For these strips, the free energy has the form \cite{jz,sdg}
$f(sq,(L_y)_{sd},q,v) = (1/L_y) \ln \lambda_{sq,(L_y)_{sd}}$, where
$\lambda_{sq,(L_y)_{sd}}$ are maximal roots of algebraic equationa of degree 5
or higher. Hence, it is thus not possible to obtain a closed-form analytic
solution for this root.  We thus follow the same high-precision numerical
procedure as described above.

\begin{figure}[hbtp]
\centering
\leavevmode
\epsfxsize=2.2in
\begin{center}
\leavevmode
\epsffile{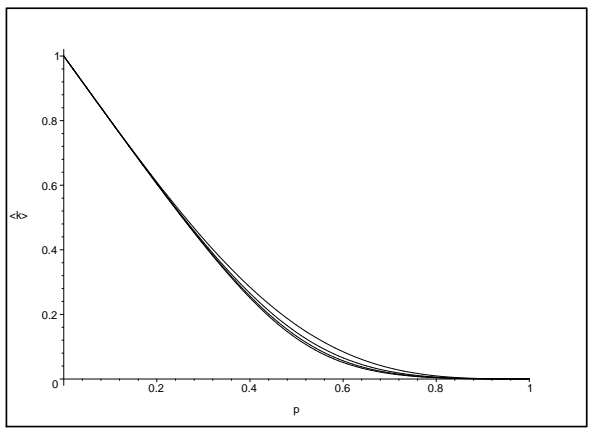}
\end{center}
\vspace{-10mm}
\caption{\footnotesize{Plots of $\langle k \rangle$, as a
function of $p \in [0,1]$, for infinite-length, finite-width
self-dual strips of the square lattice. For a given $p$, in order
of descending value of $\langle k \rangle$, the curves refer to are for 
$1_{sd} \le (L_y)_{sd} \le 4_{sd}$.}}
\label{sd}
\end{figure}

\section{Strips of the Triangular Lattice}

\subsection{Free Transverse Boundary Conditions}

\subsubsection{$2_F$} 

\begin{figure}[hbtp]
\centering
\leavevmode
\epsfxsize=2.2in
\begin{center}
\leavevmode
\epsffile{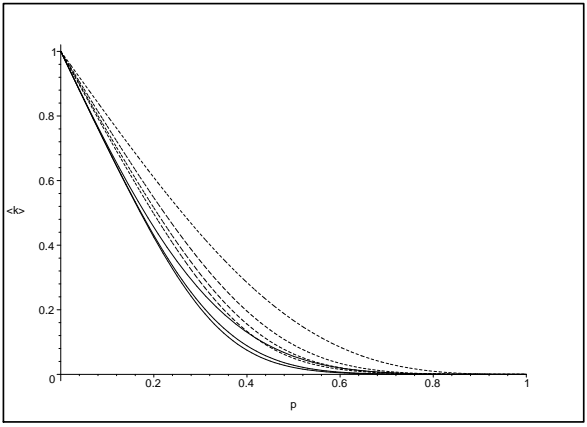}
\end{center}
\vspace{-10mm}
\caption{\footnotesize{Plots of $\langle k \rangle$, as a
function of $p \in [0,1]$, for infinite-length, finite-width
strips of the triangular lattice. The dashed and solid curves
refer to free and periodic transverse boundary conditions, respectively. 
For a given $p$, the
dashed curves are, in order of descending value of $\langle k \rangle$, for
$2_F \le (L_y)_F \le 5_F$, and the solid curves are, in the same order, for
$2_P \le (L_y)_P \le 4_P$.}}
\label{tri}
\end{figure}

The free energy for the Potts model on this strip is \cite{ta}
\beq
f(tri,2_F,q,v) = \frac{1}{2} \ln \lambda_{t2F}
\label{ftri}
\eeq
where
\beq
\lambda_{t2F}=\frac{1}{2}\biggl [ T_{t2F} + (3v+v^2+q)\sqrt{R_{t2F}} \
\biggr ]
\label{lamtri}
\eeq
with
\beq
T_{t2F}=v^4+4v^3+7v^2+4qv+q^2
\label{t12tri}
\eeq
and
\beq
R_{t2F} = q^2+2qv-2qv^2+5v^2+2v^3+v^4 \ . 
\label{rs12tri}
\eeq
 From this we calculate
\beq
\langle k \rangle_{tri,2_F} = \frac{(1-p)^3}{1-p+p^2} \ . 
\label{ktxy2}
\eeq
Note that this expression for $\langle k \rangle$ is the same as that for the
$L_y=1$ self-dual strip in eq. (\ref{ksdg}).  This provides an illustration of
the fact that two different families of lattice strips may have the same
average cluster number $\langle k \rangle$.  We plot this cluster number
$\langle k \rangle$ in Fig. \ref{tri}, together with the cluster numbers for
the various strips of the triangular lattice with greater widths and free or
periodic boundary conditions.  The values of $\langle k \rangle$ for
$p=p_{c,tri}$ are listed in Table \ref{kcvalues}.  The Taylor series expansions
of eq. (\ref{ktxy2}) for $p \to 0$ and $r=1-p \to 0$ are the same as those of
(\ref{ksdg}).

\subsubsection{$3_F,4_F,5_F$}

The free energy $f$ for the strips of the triangular lattice with width
$L_y=3,4,5$ and free transverse boundary conditions were computed in Ref. 
\cite{tt} (see also \cite{zt}).  We have used these exact analytic expressions
to obtain high-precision numerical computations of $\langle k \rangle$ for
these strips.  

\subsection{Periodic Transverse Boundary Conditions}

\subsubsection{$2_P$} 

Having explained our calculational method above for the square-lattice and
previous triangular-lattice strips, we omit the details for other lattice
strips except where we have carried out new calculations of Potts model free
energies. The free energy $f(tri,2_P,q,v)$ was calculated in Ref. \cite{ta}.
 From it we compute
\beqs
\langle k \rangle_{tri,2_P} & = & \frac{(1-p)^4(2+2p-7p^2+4p^3-p^4+2p^5-p^6)}
{2(1-2p^2+8p^3-12p^4+8p^5-2p^6)} \cr\cr
& & 
\label{ktrixpy2}
\eeqs
This has the respective Taylor series expansions for $p \to 0$ and $p \to 1$:
\beqs
\langle k \rangle_{tri,2_P}
 & = & 1 - 3p + \frac{1}{2}p^2 + 4p^3 + \frac{9}{2}p^4
-10p^5 \cr\cr & - & 10p^6+O(p^7)
\label{ktrixpy2taylorp0}
\eeqs
\beq
\langle k \rangle_{tri,2_P} = \frac{1}{2}r^4 + 2r^6 - 2r^8 + \frac{9}{2}r^{10} 
+O(r^{11}) \ . 
\label{ktrixpy2taylorp1} 
\eeq
The poles of $\langle k \rangle$ occur at
\beqs
p & \simeq & -0.3744 \ , \quad  1.6539 \ , \cr\cr
& & \quad 0.1731 \pm 0.6306i \ , \quad 1.1872 \pm 0.6924i \ .
\label{tri2Ppoles}
\eeqs
The first two poles, lying on the real $p$ axis, are closest to the points
$p=0$ and $p=1$ and determine the radii of convergence of the
series about these points to be approximately 0.3744 and 0.6539, respectively. 

Our procedure for strips with greater widths $L_y \ge 3$ is as before for the
square lattice and $(L_y)_F$ triangular-lattice strips.  The small-$p$ series
for the $tri,3_P$ lattice is
\beq
\langle k \rangle_{tri,3_P} = 1 - 3p + \frac{7}{3}p^3 + 6p^4 + O(p^5) \ . 
\label{txpy3taylorp0}
\eeq

\section{Strips of the Honeycomb Lattice}

The free energy $f(hc,2_F,q,v)$ was calculated in Ref. \cite{hca}.  From it we
obtain 
\beq
\langle k \rangle_{hc,2_F}
 = \frac{(1-p)^2(4+3p+2p^2+p^3-4p^4)}{4(1-p^4+p^5)} \ . 
\label{khc}
\eeq
This has the respective Taylor series expansions
\beq
\langle k \rangle_{hc,2_F}  = 1 - \frac{5}{4}p + \frac{1}{4}p^6 + 
\frac{1}{4}p^{10} - \frac{1}{4}p^{11} + O(p^{14})
\label{khctaylor}
\eeq
\beq
\langle k \rangle_{hc,2_F} = \frac{3}{2}r^2 + 3r^3 + O(r^4) \ . 
\label{khcrtaylor}
\eeq
The expression (\ref{khc}) has poles at
\beq
p \simeq -0.8567 \ , \quad -0.15005 \pm 0.8975i \ , \quad 
1.0784 \pm 0.4969i \ . 
\label{hc2Fpoles}
\eeq
The first of these is the nearest to the point $p=0$, so that the small-$p$
Taylor series converges for $|p| \lsim 0.8567$.  The last pair of
complex-conjugate poles is closest to $p=1$, so that the series for $r \to 0$
converges for $|1-p| \lsim 0.5031$.

\begin{figure}[hbtp]
\centering
\leavevmode
\epsfxsize=2.2in
\begin{center}
\leavevmode
\epsffile{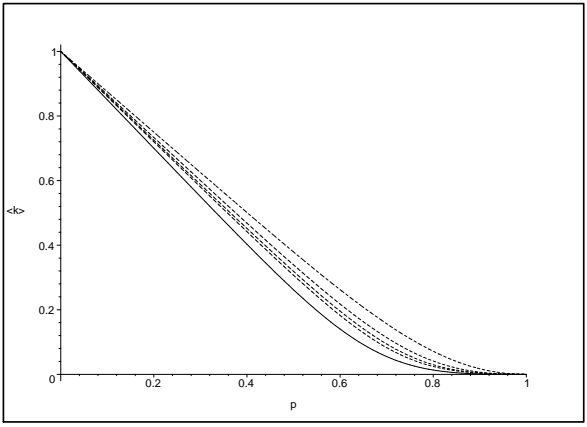}
\end{center}
\vspace{-10mm}
\caption{\footnotesize{Plots of $\langle k \rangle$, as a function of $p \in
[0,1]$, for infinite-length, finite-width strips of the honeycomb lattice.  The
dashed and solid curves refer to free and periodic transverse boundary
conditions, respectively.  For a given $p$, the dashed curves are, in order of
descending value of $\langle k \rangle$, for $2_F \le (L_y)_F \le 5_F$, and the
solid curve is for $4_P$.}}
\label{hc}
\end{figure}

Strips of the honeycomb lattice with other widths and boundary conditions are
analyzed using the same techniques as discussed above.  The resulting cluster
numbers $\langle k \rangle$ are plotted in Fig. \ref{hc} and the values for
$p=p_{c,hc}$ are listed in Table \ref{kcvalues}.

\section{Strips of the Kagom\'e Lattice}

\subsection{$2_F$}

\begin{figure}[hbtp]
\centering
\leavevmode
\epsfxsize=2.2in
\begin{center}
\leavevmode
\epsffile{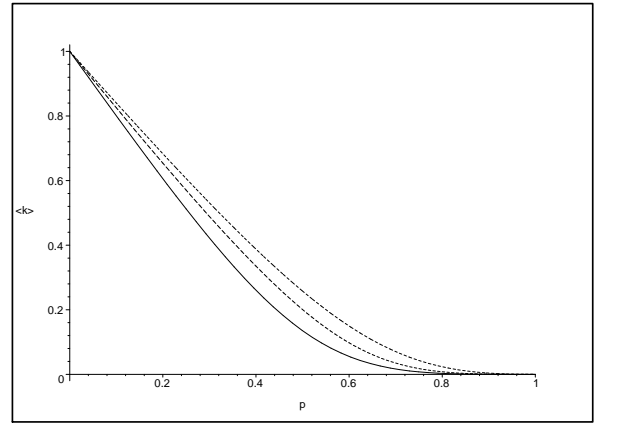}
\end{center}
\vspace{-10mm}
\caption{\footnotesize{Plots of $\langle k \rangle$, as a function of $p \in
[0,1]$, for infinite-length, finite-width strips of the kagom\'e lattice.  The
dashed and solid curves refer to free and periodic transverse boundary
conditions, respectively.  For a given $p$, the dashed curves are, in order of
descending value of $\langle k \rangle$, for $2_F \le (L_y)_F \le 3_F$, and the
solid curve is for $2_P$.}}
\label{kagfig}
\end{figure}

For the purpose of obtaining $\langle k \rangle$, we have carried out a
calculation of the free energy of the Potts model on the $2_F$ strip of the
kagom\'e lattice.  The results are sufficiently lengthy that we list them in
the appendix.  From these we calculate
\beq
\langle k \rangle_{kag,2_F} = \frac{N_{k2F}}{D_{k2F}}
\label{kkag2F}
\eeq
where
\beqs
& & N_{k2F} = (1-p)^2(5+2p-p^2-2p^3-8p^4-16p^5 \cr\cr
& & +43p^6-26p^7-2p^8+10p^9-3p^{10} \cr\cr
& & -2p^{11}+p^{12})
\label{nkag2F}
\eeqs
and
\beq
D_{k2F} = 5(1-p^4-2p^5+10p^6-10p^7+3p^8) \ . 
\label{dkag2F}
\eeq
We plot $\langle k \rangle_{kag,2_F}$ in Fig. \ref{kagfig}. 
At the value $p=p_{c,kag}$, $\langle k \rangle_{kag,2_F}$ has the approximate
value 0.22918.  The $\langle k \rangle$ for this $2_F$ Kagom\'e strip has the
following expansions in the vicinity of $p=0$ and $p=1$:
\beq
\langle k \rangle_{kag,2_F} = 1 - \frac{8}{5}p + \frac{2}{5}p^3 + \frac{1}{5}p^6
+O(p^7)
\label{kkagtaylorp0}
\eeq
\beq
\langle k \rangle_{kag,2_F} = \frac{1}{5}r^2 + \frac{8}{5}r^3 + \frac{11}{5}r^4 - 
\frac{4}{5}r^5 + O(r^6) \ . 
\label{kkagtaylorp1}
\eeq
The cluster number (\ref{kkag2F}) has poles at 
\beqs
p & \simeq & -0.5470 \pm 0.2862i \ , \quad -0.0363 \pm 0.6583i \ , \cr\cr
& & 0.7772 \pm 0.5605i \ , \quad 1.4728 \pm 0.1486i 
\label{kag2Fpoles}
\eeqs
Of these, the first and last complex-conjugate pairs are closest to $p=0$ and
$p=1$, respectively, and determine the radii of convergence of the Taylor
series expansions about these points to be approximately 0.6174 and 0.4956. 

We have also calculated $\langle k \rangle$ for the $3_F$ strip of the kagom\'e
lattice; this is plotted in Fig. \ref{kagfig}.

\subsection{$2_P$}

For the $2_P$ strip of the kagom\'e lattice we find
\beq
\langle k \rangle_{kag,2_P} = \frac{N_{k2P}}{D_{k2P}}
\label{kag2P}
\eeq
where
\beqs
& & N_{k2P} = (1-p)^4(6+12p+12p^2+4p^3-25p^4-108p^5 \cr\cr 
& & +16p^6+472p^7-706p^8+320p^9+286p^{10}-352p^{11} \cr\cr 
& & -194p^{12}+360p^{13}+120p^{14}-340p^{15}+65p^{16}+136p^{17} \cr\cr 
& & -96p^{18}+24p^{19}-2p^{20})
\label{nkag2P}
\eeqs
and
\beqs
& & D_{k2P} = 6(1-2p^4-8p^5+32p^6+40p^7-268p^8 \cr\cr 
& & +424p^9-320p^{10}+120p^{11}-18p^{12}) \ . 
\label{dkag2P}
\eeqs
A plot is given in Fig. \ref{kagfig}.  The cluster number $\langle k
\rangle_{kag,2_P}$ has poles at
\beqs
& & p \simeq -0.4660 \ , \quad 1.6556 \ , \quad -0.3443 \pm 0.2919i \ , \cr\cr
& & -0.0057 \pm 0.4751i \ , \quad 0.5325 \pm 0.48455i \ , \cr\cr
& & 1.0776 \pm 0.4384i \ , \quad 1.4785 \pm 0.2140i 
\label{kag2p_poles}
\eeqs
The first complex-conjugate pair is the nearest to the origin and sets the
radius of convergence of the small-$p$ Taylor series expansion of $\langle k
\rangle_{kag,2_P}$ as 0.4514, while the second-to-last complex-conjugate pair
is closest to the point $p=1$ and determines the radius of convergence of the
series expansion about this point to be 0.4453, to the stated accuracy.

To our knowledge, it is not known what the value of $\langle k \rangle$ is for
the (infinite) kagom\'e lattice at the numerically determined critical
percolation probability $p_{c,kag}$.  Assuming that, for a given set of
transverse boundary conditions and a given $p \in (0,1)$, $\langle k \rangle$
is a monotonically decreasing function of the strip width $L_y$ for this
lattice, as we find for other lattice strips, our results yield the upper bound
$\langle k \rangle_{kag} < \langle k \rangle_{kag,2_P} \simeq 0.11149$ at the
value $p=p_{c,kag}$ given above.  Here we use the result for the $2_P$ strip
since it is lower than the result for the $2_F$ and $3_F$ strips.

\begin{table}
\caption{\footnotesize{Values of $\langle k \rangle$ on infinite-length strips
of the lattices $\Lambda$ (where $sq,tri,hc,kag$ denote square, triangular,
honeycomb, and kagom\'e) of finite width $L_y$ at $p=p_{c,\Lambda}$, 
The transverse boundary conditions ($BC_y$) are $F$ and
$P$ for free and periodic, respectively.  We also include results for the
self-dual (sd) strips of the square lattice. The effective coordination number
$\kappa_{eff}$ is defined in eq. (\ref{kappa_eff}).  The $L_y=\infty$ values of
$\langle k \rangle_{p=p_{c,\Lambda}}$ are the values for the two-dimensional
lattice $\Lambda$, and the dashes for these entries indicate that they do not
depend on $BC_y$.}}
\begin{center}
\begin{tabular}{|c|c|c|c|c|} \hline\hline
$\Lambda$ & $BC_y$ & $L_y$ & $\kappa_{eff}$ & 
$\langle k \rangle_{p=p_{c,\Lambda}}$ \\ \hline 
sq  & F   & 1           & 2     & 0.50000   \\ \hline
sq  & F   & 2           & 3     & 0.28571   \\ \hline
sq  & F   & 3           & 3.33  & 0.21940   \\ \hline
sq  & F   & 4           & 3.50  & 0.18753   \\ \hline
sq  & F   & 5           & 3.60  & 0.16887   \\ \hline\hline
sq  & P   & 2           & 4     & 0.20000   \\ \hline
sq  & P   & 3           & 4     & 0.14103   \\ \hline
sq  & P   & 4           & 4     & 0.12150   \\ \hline
sq  & P   & 5           & 4     & 0.11284   \\ \hline\hline
sq  & sd  & 1           & 4     & 0.16667   \\ \hline
sq  & sd  & 2           & 4     & 0.14407   \\ \hline
sq  & sd  & 3           & 4     & 0.132545  \\ \hline
sq  & sd  & 4           & 4     & 0.12561   \\ \hline\hline
sq  & $-$ & $\infty$    & 4     & 0.09808   \\ \hline\hline
tri & F   & 2           & 4     & 0.35958   \\ \hline
tri & F   & 3           & 4.67  & 0.27149   \\ \hline
tri & F   & 4           & 5     & 0.22946   \\ \hline
tri & F   & 5           & 5.20  & 0.20491   \\ \hline\hline
tri & P   & 2           & 6     & 0.19091   \\ \hline
tri & P   & 3           & 6     & 0.14665   \\ \hline
tri & P   & 4           & 6     & 0.13138   \\ \hline\hline
tri & $-$ & $\infty$    & 6     & 0.11184   \\ \hline\hline
hc  & F   & 2           & 2.50  & 0.20475   \\ \hline
hc  & F   & 3           & 2.67  & 0.16000   \\ \hline
hc  & F   & 4           & 2.75  & 0.13834   \\ \hline
hc  & F   & 5           & 2.80  & 0.12560   \\ \hline\hline
hc  & P   & 4           & 3     & 0.08983   \\ \hline\hline
hc  & $-$ & $\infty$    & 3     & 0.07687   \\ \hline\hline
kag & F   & 2           & 3.2   & 0.22918   \\ \hline
kag & F   & 3           & 3.5   & 0.17220   \\ \hline\hline
kag & P   & 2           & 4     & 0.11149   \\ \hline\hline
\end{tabular}
\end{center}
\label{kcvalues}
\end{table}

\section{Discussion}

We first introduce a notion of effective coordination number.  For a graph $G$
the degree of a vertex is the number of bonds connected to this vertex. A
$\kappa$-regular graph is a graph in which all of the vertices have the same
degree, $\kappa$.  Whether a given lattice strip graph is $\kappa$-regular
depends on the longitudinal and transverse boundary conditions; for example, it
is $\kappa$ regular if one uses toroidal (doubly periodic) boundary conditions.
In the limit $L_x \to \infty$, since the longitudinal boundary conditions do
not affect the free energy $f(\{G\},q,v)$, we need only consider the effect of
the transverse boundary conditions.  The effective coordination number is
\beq
\kappa_{eff}(\{G\}) = \lim_{n \to \infty} \frac{2e(G)}{n(G)} \ . 
\label{kappa_eff}
\eeq
Clearly $\kappa_{eff}=\kappa$ for a regular lattice. 
For regular lattice strips with periodic transverse boundary conditions, the
value of $\kappa_{eff}$ is the same as the value for the corresponding
two-dimensional lattice.  For strips with free transverse boundary conditions,
we have 
\beq
\kappa_{eff}(\Lambda,(L_y)_F) = \kappa_\Lambda 
\biggl ( 1 - \frac{\alpha}{L_y} \biggr )
\label{kappa_L}
\eeq
where $\kappa_\Lambda=4,6,3$ for $\Lambda=sq,tri,hc$ and 
\beq
\alpha_{sq}=\frac{1}{2} \ , \alpha_{tri} = \frac{2}{3} \ , 
\quad \alpha_{hc} = \frac{1}{3} \ .
\label{alpha}
\eeq
For the cyclic self-dual strips of the square lattice, the single external
vertex connected to each of the sites on one side of the strip has a degree
$L_x$ that diverges as $L_x \to \infty$.  The $L_x(L_y-1)$ interior vertices
have degree 4, while the $L_x$ vertices on the rim have degree 3.  Together,
these lead, in the limit $L_x \to \infty$, to the result $\kappa_{sq,sd} = 4$.
Finally, for the kagom\'e strips with free transverse boundary conditions
\beq
\kappa_{eff}(kag,(L_y)_F) = 4\Bigl ( 1 - \frac{1}{3L_y-1} \Bigr ) 
\label{kappakag}
\eeq
while for the kagom\'e strips with periodic transverse boundary conditions, 
$\kappa_{eff}=4$, the same value as for the infinite two-dimensional kagom\'e
lattice.

 From our calculations we find a number of generic features:

\begin{itemize}

\item 
We have shown that $\langle k \rangle$ is a (real) analytic function of $p$ in
the interval $0 \le p < 1$.  At the critical percolation probability $p=1$ for
these quasi-1D strips, our exact results for $\langle k \rangle$ are also
analytic, although some other quantities in percolation, such as the
percolation probability $P(p)$ and the cluster size $S(p)$ are not, as is
evident from the well-known 1D case.

\item 
As the curves in the figures show, with an increase in strip width
$L_y$, $\langle k \rangle$ is consistent with approaching a limiting function
of $p$.  This is in accord with one's expectation.  

\item 
For a given $p$ in the interval between 0 and 1, and for a given type of
lattice strip, as the width $L_y$ increases, $\langle k \rangle$ decreases, so
that the approach to the asymptotic value for the 2D lattice is from above, in
the cases that we have computed.  For strips with free transverse boundary
conditions, increasing $L_y$ increases $\kappa_{eff}$, so the decrease of
$\langle k \rangle$ is associated with an increase in the effective
coordination number.  This is reasonable, since, heuristically, for a fixed
value of $p$, there is a greater probability of having a percolating cluster on
a lattice of higher coordination number, so that more sites are part of this
cluster and there are fewer separate clusters per site.  This is also reflected
in the monotonic decrease of $p_{c,\Lambda}$ with increasing $\kappa_\Lambda$
for most higher-dimensional lattices.  (However, we recall that counterexamples
to this general monotonic decrease of $\langle k \rangle$ with increasing
coordination number are known \cite{marck,counter}.)

For strips with periodic transverse boundary conditions,
the decrease of $\langle k \rangle$ at a fixed $p$ with increasing width $L_y$
is not associated with an increase in $\kappa_{eff}$, since $\kappa_{eff}$ is
constant for these strips (and equal to the two-dimensional value); here one
may interpret the decrease as being simply due to a reduction in the
finite-width effects that enables the percolation quantities to approach their
two-dimensional values.  

\item
For a given lattice type, we find some examples where the curve for $\langle k
\rangle$ calculated on a strip of width $L_y$ with periodic transverse boundary
conditions will cross the curve for $\langle k \rangle$ for the same lattice
and a different $L_y$ and free transverse boundary conditions.  For example, as
is evident in Fig. 1, the curve for $\langle k \rangle$ on the $sq,2_P$ strip
lies below those for $\langle k \rangle$ on the $sq,(L_y)_F$ strips at small
$p$, but sequentially crosses the latter as $p$ increases and lies above them
(except for $L_y=1,2$) as $p \to 1^-$.  Similar behavior is observed, e.g., on
the strips of the triangular lattice.  These also constitute examples of how
$\langle k \rangle$ calculated on a strip with a larger value of $\kappa_{eff}$
than that of another strip can be larger than $\langle k \rangle$ for the
latter strip.  For instance, $\kappa_{eff}=\kappa=4$ for the $sq,2_P$ strip,
which is larger than the value $\kappa_{eff}=3.6$ for the $sq,5_F$ strip;
however, $\langle k \rangle$ on the former strip is larger than $\langle k
\rangle$ on the latter for $p \gsim 0.36$.  This dependence on transverse
boundary conditions is consistent with disappearing as the strip width $L_y \to
\infty$, consistent with the approach to a single limiting function $\langle k
\rangle$ for the corresponding 2D lattice.  Although we have not proved
rigorously that the function $\langle k \rangle$ obtained via this limiting 
sequence (taking $L_x \to \infty$ first and then taking $L_y \to \infty$) 
is identical to the function $\langle k \rangle$ obtained via the usual
two-dimensional thermodynamic limit ($L_x \to \infty$, $L_y \to \infty$ with
$L_y/L_x$ a nonzero finite number), this conclusion is consistent with our
findings. 

\item

We have used the values of $\langle k \rangle$ at $p=p_{c,\Lambda}$ as a
measure of how rapidly, for a given $p$, the cluster number calculated on
infinite-length, finite-width strips approaches the value for the
two-dimensional lattice.  These values are listed in Table \ref{kcvalues}.
Even for the modest strip widths considered here, one sees that (i) these
values approach the known values of $\langle k \rangle$ on the corresponding
two-dimensional lattices reasonably quickly, and (ii) this approach is more
rapid when one uses periodic transverse boundary conditions, as is expected,
since the latter minimize finite-width effects.  For example, for the strip of
the square lattice with $L_y=5$ and periodic transverse boundary conditions,
$\langle k \rangle$ evaluated at $p=p_{c,sq}$ is about 15 \% larger than the
value (\ref{kcsq}) for the square lattice, while $\langle k \rangle$ for the
$tri,4_P$ and $hc,4_P$ strips, evaluated at the respective $p_{c,tri}$ and
$p_{c,hc}$, are both about 17 \% larger than the corresponding values 
(\ref{kctri}) and (\ref{kchc}) for the triangular and honeycomb lattices. 

\item 

We find that for these strips, the small-$p$ series expansions
of $\langle k \rangle$ have the leading terms 
\beq
\langle k \rangle = 1 - \biggl ( \frac{\kappa_{eff}}{2} \biggr ) p + ...
\label{kp0linear}
\eeq
which are analogous to the structure that these series have for regular
lattices of dimension $d \ge 2$.  Higher-order terms in the series for the
strips of small widths are not expected to coincide with those in the series for
the two-dimensional lattices, and one sees that they do not.

\item 

An interesting output of our analysis is the exact determination, for various
infinite-length, finite-width strips, of the singularities of $\langle k
\rangle$ in the complex $p$ plane.  As we have shown, for many strips these
(real and/or complex) singularities outside the physical interval $[0,1]$ occur
sufficiently close to the points $p=0$ and $p=1$ that they render the radii of
convergence of the respective Taylor series expansions about these points less
than unity, although the actual functions $\langle k \rangle$ themselves are
analytic functions on $p \in [0,1]$.  Although the strip widths are probably
too small to justify a detailed comparison with unphysical singularities for
percolation quantities in two dimensions, this generic property - the presence
of unphysical singularities that determine the radii of the Taylor series
expansions about the points $p=0$ and $p=1$ to be less than $p_c$ for the given
type of lattice - is similar to what was found in analyses of series for the
percolation problem on two-dimensional lattices \cite{sykesseries} (and
three-dimensional lattices \cite{sykes3D}).

\item 

Finally, we have discussed how, for a given infinite-length, finite-width
strip, the unphysical singularities have a connection with the locus ${\cal
B}_{qn}$, which is the continuous accumulation set of the zeros of the Potts
model partition function in the $p$ (or equivalently the $v$) plane obtained by
first letting $n \to \infty$ and then $q \to 1$.  In particular, we find that
these unphysical singularities lie on ${\cal B}_{qn}$. The noncommutativity of
eq. (\ref{bnoncomm}) analyzed in Ref. \cite{a} plays a crucial role here, since
${\cal B}_{nq}$, obtained by first letting $q \to 1$ and then $n \to \infty$,
is trivial.  Our results motivate further study on this topic.

\end{itemize}

\section{Conclusions} 

In summary, we have presented exact calculations of the average cluster number
per site $\langle k \rangle$ for the bond percolation problem on
infinite-length, finite-width strips of the square, triangular, honeycomb, and
kagom\'e lattices, with both free and periodic transverse boundary conditions.
We believe that these results are a useful extension beyond the one-dimensional
result toward two dimensions and provide insight into the form of $\langle k
\rangle$ as a function of the bond occupation probability $p$.

\begin{acknowledgements}

We thank R. Ziff for helpful comments.  The research of R.S. was partially
supported by the grant NSF-PHY-00-98527.

\end{acknowledgements}

\section{Appendix}

In this appendix we give the free energy for the Potts model on the $2_F$ 
strip of the kagom\'e lattice.  We find 
\beq
f(kag,2_F,q,v) = \frac{1}{5} \ln \lambda_{k2F}
\label{fkag2}
\eeq
where
\beq
\lambda_{k2F}=\frac{1}{2}[ T_{k2F} + \sqrt{R_{k2F}} \ ]
\label{lamkag}
\eeq
with
\beqs
& & T_{k2F} = v^8+8v^7+v^6q+29v^6+20v^5q+10v^4q^2 \cr\cr
& + & 2v^3q^3+42v^5+61v^4q+54v^3q^2+28v^2q^3 \cr\cr
& + & 8vq^4+q^5
\label{Tkag2F}
\eeqs
and
\beqs
& & R_{k2F} =  v^{16}+16v^{15}+2v^{14}q+114v^{14}+32v^{13}q \cr\cr
& & -3v^{12}q^2-4v^{11}q^3+484v^{13}+288v^{12}q+52v^{11}q^2 \cr\cr
& & -28v^{10}q^3-12v^9q^4-2v^8q^5+1329v^{12}+1572v^{11}q \cr\cr
& & +1098v^{10}q^2+520v^9q^3+192v^8q^4+48v^7q^5+6v^6q^6 \cr\cr
& & +2196v^{11}+4350v^{10}q+5196v^9q^2+4344v^8q^3 \cr\cr
& & +2628v^7q^4+1114v^6q^5+312v^5q^6+52v^4q^7+4v^3q^8 \cr\cr
& & +1620v^{10}+4572v^9q+7413v^8q^2+8284v^7q^3+6732v^6q^4 \cr\cr
& & +4028v^5q^5+1766v^4q^6+556v^3q^7+120v^2q^8+16vq^9 \cr\cr
& & +q^{10} \ . 
\label{Rkag2F}
\eeqs
(It can be checked that the $v=-1$ special case of this $f$ coincides with the
degeneracy per site, $W$, in Refs. \cite{kagref}.)


\begin{thebibliography}{99}
\bibitem{sqpxy} 
For periodic longitudinal boundary condition, the partition
function for the $sq,2_F$ strip involves six $\lambda$'s \cite{a}, but since the
free energy is independent of the longitudinal boundary conditions in the case
$L_x \to \infty$ of interest, it will suffice to consider only the case of free
longitudinal boundary conditions.
\end{thebibliography}
\end{document}